\documentclass[pra,aps,twocolumn,showpacs,floatfix]{revtex4-1}
\usepackage{amssymb}
\usepackage{amsfonts}
\usepackage{hyperref}
\usepackage{graphicx}
\usepackage{dcolumn}
\usepackage{amsmath}
\usepackage{bm,amsmath,verbatim}
\usepackage{mathrsfs}
\usepackage{color}
\usepackage[noend]{algpseudocode}
\usepackage{algorithmicx,algorithm}
\usepackage{dsfont}
\usepackage{booktabs}
\usepackage{multirow}
\usepackage{makecell}
\usepackage{braket}

\begin{document}
\title{Hinge solitons in three-dimensional second-order topological insulators}
\author{Yu-Liang Tao$^{1}$}
\author{Ning Dai$^{1}$}
\author{Yan-Bin Yang$^{1}$}
\author{Qi-Bo Zeng$^{1}$}
\author{Yong Xu$^{1,2}$}
\email{yongxuphy@tsinghua.edu.cn}
\affiliation{$^{1}$Center for Quantum Information, IIIS, Tsinghua University, Beijing 100084, People's Republic of China}
\affiliation{$^{2}$Shanghai Qi Zhi Institute, Shanghai 200030, People's Republic of China}

\begin{abstract}
Higher-order topological insulators have recently witnessed rapid progress in various fields ranging from condensed matter
physics to electric circuits. A well-known higher-order state is the second-order topological insulator in three dimensions with gapless states localized on the hinges. A natural question in the context of nonlinearity is whether solitons can exist
on the hinges in a second-order topological insulator.
Here we theoretically demonstrate the existence of stable solitons localized on the hinges of a second-order topological insulator in three dimensions when nonlinearity is involved. By means of systematic numerical study, we find that the soliton has strong localization in real space and propagates along the hinge unidirectionally without changing its shape.
We further construct an electric network to simulate the second-order topological insulator. When a nonlinear inductor
is appropriately involved, we find that the system can support a bright soliton for the voltage distribution
demonstrated by stable time evolution of a voltage pulse.
\end{abstract}
\maketitle

\section{INTRODUCTION}
Solitons, solitary waves travelling without changing their shapes, result from the balance
between dispersion and nonlinearity. Solitons exist in various nonlinear systems, such as
nonlinear optics~\cite{YSKivsharbook,TDauxoisbook,Zeng2020commun}, Bose-Einstein condensates (BECs)~\cite{AShabat1972SPJ,SBurger,PGKevrekidibook,KartashovPRL20192,Zeng2019Adv} and Fermi superfluids~\cite{JDziarmaga2005,Antezza2007PRA,TYefsah2013Nature,Xu2014PRL,MJHKu2014PRL,PZou2016PRl}. They usually
stably exist in a one-dimensional system. Yet, it is a challenging problem to create solitons,
especially for a bright soliton with a wave density bump, in higher-dimensional systems.
The difficulty lies in the fact that the nonlinear Schr\"{o}dinger equation for the ubiquitous
cubic local nonlinearity leads to
critical and supercritical wave collapse~\cite{CSulem1999BOOK,KartashovNatrp2019} in two and three dimensions, respectively.
To generate stable higher-dimensional bright solitons, one can consider nonlocal
nonlinearity, such as the nonlinearity resulted from the dipole-dipole interactions in
BECs~\cite{PPedri2005PRL,TLahaye2009RPP}, and modified kinetic energy, such as involving the spin-orbit coupling~\cite{YCZhang2009PRL,KartashovPRL2019}.
An alternative perspective is to engineer solitons at the boundaries with lower dimension
in a higher-dimensional system. Since topological insulators support lower dimensional
edge states, they provide an ideal platform to realize the boundary-localized solitons.
Indeed, such solitons have been theoretically predicted in a two-dimensional
topological insulator with
one-dimensional edges~\cite{Lumer2013PRL,Mark2014PRA,Ablowitz2015,Ablowitz20152,DLeykam2016PRL2,YVKartashov2016,Gulevich2017SR,Ablowitz2017PRA,DDJM2019EMl,Smirnova2019LPR,Elyasi2019PRB,Pal2019PRE,SKIvanov2020}. Remarkably, the topological edge solitons have been experimentally observed very recently~\cite{Mukherjee2020Science,Ivanov2020acs,Zhang2020natc}.
Interestingly, solitons have also been numerically found
in Weyl semimetals~\cite{CShang2018PRA}.

Traditional topological insulators in $n$ dimensions support edge states localized at $(n-1)$-dimensional boundaries,
such as the Chern insulator. Recently, topological phases have been generalized to
the higher-order case where gapless edge states are localized at $(n-m)$-dimensional (with $m>1$) boundaries
~\cite{Fritz2012PRL,ZhangFan2013PRL,Slager2015PRB,Taylor2017Science,Brouwer2017PRL,FangChen2017PRL,Wan2017arXiv,Taylor2017PRB,Ezawa2018PRL,WangZhong2018PRL,Fulga2018PRB,Watanabe2018PRB,You2018PRB,Neupert2018NP,Taylor2018PRB2,Brouwer2018PRB,ZhangFan2018PRL,Huber2018Nature,
Bahl2018Nature,SImhof2018NP,Brouwer2019PRX,Lee2019PRL,Roy2019PRB,Fengliu2019PRL,SYang2019PRL,
YanBin2019Arxiv1,Schindler2018SA,Kai2020prl,Vincent2020prl,Qibo2020prb}.
Such insulators are coined higher-order topological insulators, or specifically, $m$th-order topological insulators.
For instance, in two dimensions (2D), there exists
type-I~\cite{Taylor2017Science} and type-II~\cite{YanBin2019Arxiv1} quadrupole topological
insulators that support zero-energy modes localized at the corners. In three dimensions (3D),
there exists a second-order topological insulator hosting chiral modes localized on the hinges,
when an appropriate term breaking the time-reversal symmetry is involved in a $\mathbb{Z}_2$ topological insulator~\cite{Schindler2018SA}. In the context of nonlinearity, it is natural to ask whether the second-order topological system
can support solitons localized on the hinges when nonlinearity is involved.

Several higher-order topological phases have been experimentally observed in several systems~\cite{Huber2018Nature,Bahl2018Nature,Neupert2018NP,SImhof2018NP}.
One of them uses electric circuits~\cite{SImhof2018NP} which have been demonstrated to be a powerful platform to simulate various topological phases
in lattice models, such
as the Su-Schrieffer-Heeger model~\cite{Lee2018CP}, topological nodal semimetals~\cite{YHLu2019PRB,KLuo2018R}, topological amorphous metals~\cite{YBY2019PRL}, non-Hermitian Aubry-Andr\'e-Harper models~\cite{QBZ2020PRB,Jiang2019PRB}, two-dimensional higher-order topological phases~\cite{SImhof2018NP,YanBin2019Arxiv1,Qibo2020prb} and others~\cite{Ningyuan2015PRX,Albert2015PRL,Luo2018arXiv,Zhu2019PRB,ZQZhang2019PRB,Haenel2019PRB,Liu2020PR,Olekhno2020NC,MEzawa2019PRB,Hofmann2019PRL}. However, whether the electric circuits can be used to realize the 3D second-order topological insulator
has not been explored. In addition,
an electric network becomes nonlinear when either nonlinear capacitors or inductors (such as the nonlinear transmission lines) are involved, leading to electric solitons in the form of voltage waves~\cite{Suzuki1970PSJ,Suzuki1973PIE,WMLiubook}. One may wonder whether the electric circuit can support
hinge solitons in the 3D second-order topological insulator.

In this paper, we theoretically predict the existence of solitons in a 3D second-order topological insulator
when nonlinearity is involved. Such solitons result from the balance between nonlinearity and dispersion of
the hinge modes. For the higher-order topological insulator, there are two regimes: strong regime
with chiral hinge modes and weak regime with hinge modes that are not chiral. We find that the solitons can exist
in both of these two regimes. Yet, we show that the soliton in the strong regime is more stable than the one in the weak regime, which gradually slows down. Furthermore, we propose a scheme to realize the second-order
topological insulator in electric circuits with nonlinearity that can be realized by a voltage-controlled variable
inductor. By simulating the dynamics of the circuit, we finally show that a bright soliton can stably exist in the system.

\section{Linear Model Hamiltonian}
We start by considering the model described by the following Hamiltonian in momentum space~\cite{Schindler2018SA}
\begin{align}
\label{EqHLk}
H_L(\textbf{k})=&(M+J\sum_{\nu=x,y,z}\cos k_\nu)\tau_z\sigma_0+\Delta_1\sum_{\nu=x,y,z}\sin k_\nu\tau_x \sigma_\nu \nonumber\\
&+\Delta_2(\cos k_x-\cos k_y)\tau_y \sigma_0,
\end{align}
where $\sigma_\nu$ and $\tau_\nu$ represent Pauli matrices acting on different degrees of freedom, $\sigma_0$ and $\tau_0$ refer to $2\times2$ identity matrices. $J$, $M$, $\Delta_1$ and $\Delta_2$ are real parameters. When $\Delta_2=0$,
the Hamiltonian describes the 3D $\mathbb{Z}_2$ topological insulator~\cite{Qi2008PRB}.
It respects both time-reversal symmetry, i.e., $\mathcal{T}H({\bf k})\mathcal{T}^{-1}=H(-{\bf k})$
with $\mathcal{T}=i\sigma_y \kappa$ ($\kappa$ denotes the complex conjugation operator),
and the rotational symmetry, i.e., ${C_4}^z H({\bf k}) ({{C_4}^z})^{-1}=H(-k_y,k_x,k_z)$ with
${C}^z_4 =e^{-i\frac{\pi}{4}\sigma_z}$.
When $1<|M/J|<3$, the Hamiltonian represents a strong topological
insulator with odd number of Dirac points in the energy spectra of surface states; when $|M/J|<1$, it
describes a weak
topological insulator with even number of Dirac points in the surface energy spectra~\cite{LFu2007PR,LFu2007PRL}.

To generate the higher-order topological insulating phase, Ref.~\cite{Schindler2018SA} introduces the term involving $\Delta_2$ that
breaks both the time-reversal symmetry and the rotational symmetry but preserves the symmetry of their product,
i.e., $C_4^z \mathcal{T}$. This term provides an effective mass, opening the gap of the Dirac points on the
surface vertical to either $x$ or $y$ axis. Since the signs of the Dirac masses are opposite for neighboring surfaces,
the hinges of their intersections exhibit chiral modes, similar to the chiral modes of a Chern insulator.
It has been found that four chiral modes exist on the four hinges between the surfaces normal to $x$ and $y$ axes.
The topological property of this higher-order phase can be characterized by the Chern-Simons term
\begin{align}
	\label{CS_term}
\theta=\frac{1}{4\pi}\int \mathrm{d}^3k \epsilon_{abc}\mathrm{tr}[\mathcal{A}_a \partial_b \mathcal{A}_c+i\frac{2}{3}\mathcal{A}_a \mathcal{A}_b \mathcal{A}_c],
\end{align}
where $\mathcal{A}_{a;n,n'}=i\bra{\varphi_n(\bf{k})}\partial_{k_a}\ket{\varphi_{n'}(\bf{k})}$ with $a,b,c\in \{x,y,z\}$ is the $k_a$-component of Berry connection, $\ket{\varphi_{n}(\bf{k})}$ is the Bloch eigenstate of $H_L(\textbf{k})$ and $n,n'$ are running over the occupied bands of the system. The topological invariant is quantized to $0$ or $\pi$ by $C_4^z \mathcal{T}$ symmetry~\cite{Schindler2018SA}.

\begin{figure}[t]
\includegraphics[width = 1\linewidth]{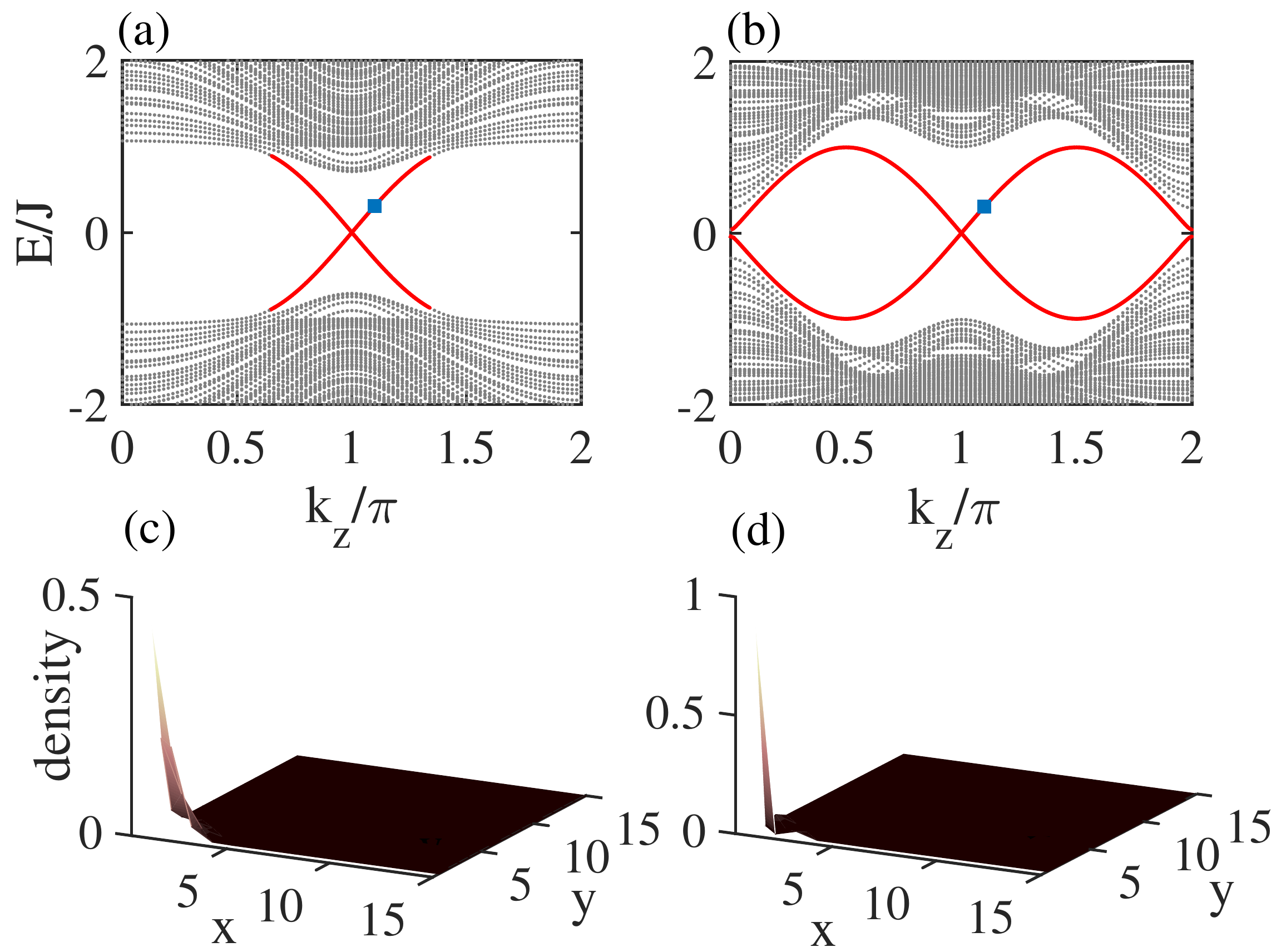}
\caption{Energy spectra with respect to $k_z$ for a system with open boundaries along $x$ and $y$
and periodic boundary along $z$ for (a) $M=2J$ and (b) $M=0.8J$, corresponding to strong and weak topological insulators, respectively, when $\Delta_2=0$.
The red lines depict the states localized on the four hinges. The hinge modes are chiral in the former case but not
in the latter. The density profiles of a hinge mode in the $(x,y)$ plane for (a) $M=2J$ and $k_z=8\pi/7$
and (b) $M=0.8J$ and $k_z=8\pi/7$, characterized by the density distribution in the $(x,y)$ plane,
$\sum_{\lambda=1,2,3,4}{|\varphi_{k_z,\lambda}(x,y)|^2}$,
where $\varphi_{k_z}$ is the eigenstate corresponding to the eigenenergy labelled by the blue squares in (a-b) of the Hamiltonian (\ref{EqHLk})
with open boundaries along the $x$ and $y$ directions and periodic boundary along the $z$ direction,
i.e., $H_L(k_z)\varphi_{k_z}=\varepsilon_{k_z}\varphi_{k_z}$.
Here $\Delta_1=\Delta_2=J$.}	
\label{Fig1}
\end{figure}

For simplicity without loss of generality, we set $\Delta_1=\Delta_2=J$. In Fig.~\ref{Fig1},
we plot the energy spectra in a geometry with open boundaries along $x$ and $y$ and periodic boundaries along $z$.
We see that there exist chiral modes localized on the hinges when $M=2J$ in the strong topological
regime. When $M=0.8J$ in the weak topological regime, we can still
observe the hinge-localized modes, but they are not chiral. We find solitons in both of these two cases, but
their stabilities manifested by time evolution are different, which will be discussed in detail in the following.

\section{Hinge solitons}
To create the hinge solitons, we add nonlinear terms into the higher-order topological system so that the dynamics is governed by the following nonlinear Schr\"odinger equation,
\begin{equation}
\label{Eq_nonSch_q}
i\hbar\frac{\partial {\psi}({\bf r},t)}{\partial t}=\left[{H}_L +H_{N}({\psi})\right]{\psi}({\bf r},t),
\end{equation}
where $H_{N}({\psi})_{({\bf r},\lambda),({\bf r}^\prime,\lambda^\prime)}=\delta_{{\bf r}{\bf r}^\prime}\delta_{\lambda \lambda^\prime}g_{\lambda}|\psi_\lambda({\bf r})|^2$ represents the short-range nonlinear interactions
with $g_\lambda$ denoting their strength for each component,
$H_L$ is the real space counterpart of the Hamiltonian~(\ref{EqHLk})
and $\psi({\bf r},t)$ is the wave function.
Here we consider a cubic local nonlinear interaction, which widely exists in ultracold atoms and
nonlinear optics~\cite{BECbook,TDauxoisbook}. For simplicity, we set $g_\lambda=g>0$.

\begin{figure*}[t]
\includegraphics[width =\textwidth]{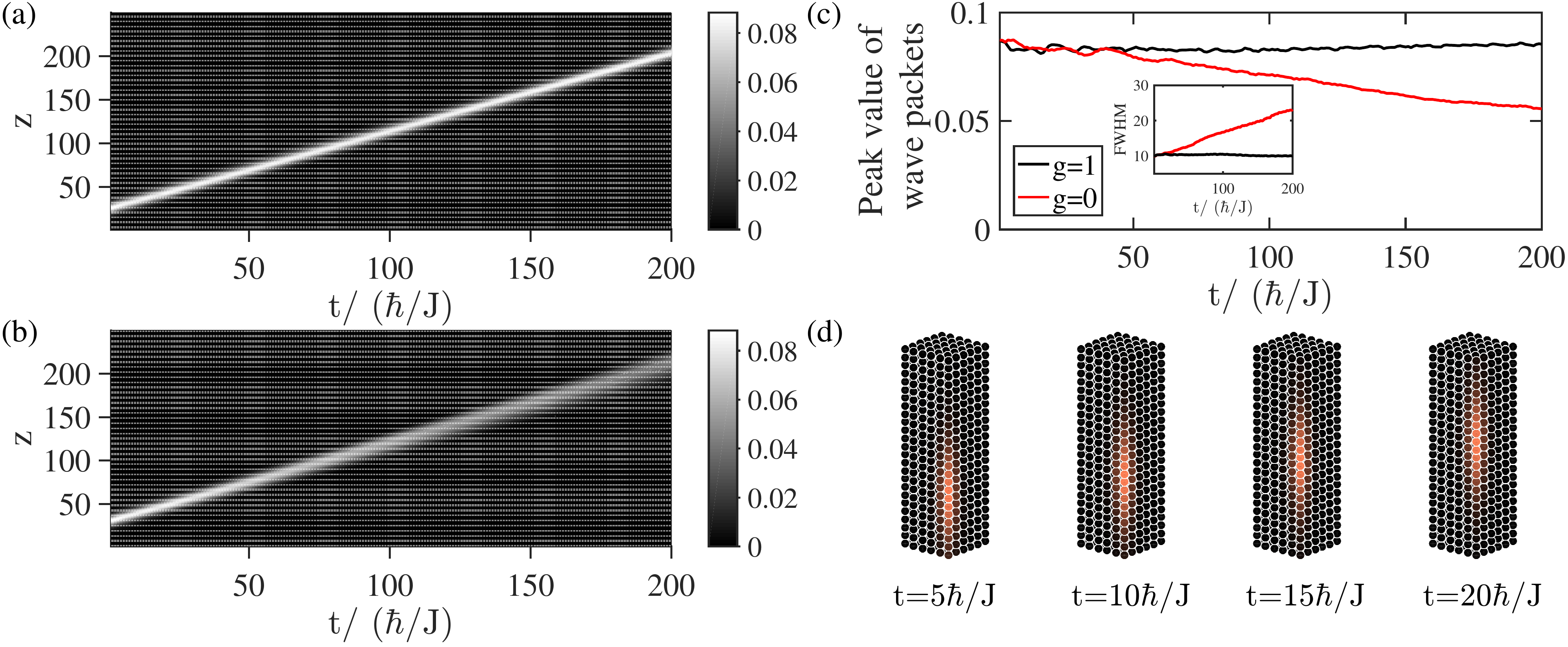}
\caption{The time evolution of a wave packet characterized by $\sum_{\lambda=1,2,3,4}|\psi_\lambda|^2$ for (a) $g=J$ and (b) $g=0$. The wave packet is initialized
at $t=0$ based on Eq.~(\ref{Eq_psi_tayl}) and Eq.~(\ref{Eq_a_bri}) with $k_z=8\pi/7$, $\varepsilon''=-0.4374$ and $\delta=0.007J$.
(c) The evolution of the peak value of wave packets over time with the black and red lines denoting the cases with $g=J$
and $g=0$, respectively. The inset plots the change of the full width at half maximum (FWHM) of the wave packets as time evolves. (d) Visualization of the propagation of the soliton over time in real space. See also the supplementary material for the dynamics of a soliton
when it enters into the top surface.
Here $M=2J$.}	
\label{Fig2}
\end{figure*}

Following Ref.~\cite{Ablowitzbook}, we construct a wave packet using the hinge modes of $H_L$ to determine
a soliton solution,
\begin{align}\label{Eq_psi_int}
{\psi}_\lambda({\bf r},t)=&\int_{-\pi}^{\pi} a(\kappa,t)\varphi_{k_z+\kappa,\lambda}(x,y)\nonumber\\
&e^{-i\varepsilon_{k_z+\kappa} t+i(k_z+\kappa)z}d\kappa,
\end{align}
where $\lambda=1,2,3,4$ is the component index, $\varphi_{k_z}$ denotes the hinge mode of $H_L(k_z)$
corresponding to the eigenenergy $\varepsilon_{k_z}$
and $a(\kappa, t)$ denotes the weights for the corresponding hinge mode over time.

We now apply the Taylor-series expansion in $\kappa$ for
$\varphi_{k_z+\kappa,\lambda}(x,y)$ in the above integral to simplify Eq.~(\ref{Eq_psi_int}) to
\begin{align}\label{Eq_psi_tayl}
{\psi}_\lambda({\bf r},t)=e^{-i\varepsilon_{k_z} t+ik_z z}\sum_{j=0}^{\infty} \frac{(-i)^j}{j!}\frac{\partial^j \varphi_{k_z,\lambda}}{\partial k_z^j}\frac{\partial^j a(z,t)}{\partial z^j},
\end{align}
where $a(z,t)=\int_{-\pi}^{\pi}a(\kappa,t)e^{i\kappa z}d\kappa$ is the envelope function of the corresponding soliton.
Applying ${H}_L$ to the wave function $\psi$ in Eq.~(\ref{Eq_psi_tayl}) and then performing the Taylor-series expansion for
both $\varepsilon_{k_z+\kappa}$ and $\varphi_{k_z+\kappa,\lambda}(x,y)$ yields
\begin{align}\label{Eq_Hpsi}
{H_L}{\psi}=e^{-i\varepsilon_{k_z} t+ik_z z}\sum_{j=0}^{\infty} \frac{(-i)^j}{j!}\frac{\partial^j (\varepsilon_{k_z} \varphi_{k_z})}{\partial k_z^j}\frac{\partial^j a(z,t)}{\partial z^j}.
\end{align}
Suppose that $\varphi_{k_z}$ varies much more slowly with respect to $k_z$ than $\varepsilon_{k_z}$,
we can make the approximation that $\partial^j(\varepsilon \varphi)/\partial k_z^j\approx \varphi\partial^j\varepsilon/\partial k_z^j$. By further assuming that $a(z,t)$ varies slowly in time and space, we obtain
an approximate nonlinear Schr\"odinger equation for the envelop function
\begin{align}\label{Eq_equ_a}
i\frac{\partial a}{\partial t}+i\varepsilon'\frac{\partial a}{\partial z}+\frac{\varepsilon ''}{2}\frac{\partial^2 a}{\partial z^2}-
g_{\textrm{eff}}|a|^2 a=0,
\end{align}
where $\varepsilon'=\partial\varepsilon_{k_z}/\partial k_z$, $\varepsilon''=\partial^2\varepsilon_{k_z}/\partial k_z^2$ and
$g_{\textrm{eff}}=g\sum_{x,y,\lambda}|\varphi_{k_z,\lambda}(x,y)|^4$.
Evidently, this equation is exactly the same as the Gross-Pitaevskii equation for a BEC in a free space in a reference moving with
the velocity of $-\varepsilon'$ with $1/\varepsilon''$ playing the role of an effective mass. If $g_{\textrm{eff}}=0$,
this equation does not support a localized solution. Because of the chiral property of the hinge modes in the strong regime,
the solitons can only move along one direction on a hinge, which is fundamentally different from solitons in a one-dimensional system
whose velocities can be either positive or negative.

When $\varepsilon''<0$ for $g_{\textrm{eff}}>0$, this equation has a bright soliton solution with a density peak,
\begin{eqnarray}\label{Eq_a_bri}
a_{{B}}(z,t)=&&(2\delta/g_{\textrm{eff}})^{1/2} \nonumber \\
&&\rm{sech}\left[(-2\delta/\varepsilon'')^{1/2}(z-\varepsilon' t)\right]e^{-i\delta t},
\end{eqnarray}
where $\delta>0$ describes the energy difference between the nonlinearity-induced energy eigenvalue $\mu$ and the linear one $\varepsilon_{k_z}$ at $k_z$.
While the Eq.~(\ref{Eq_a_bri}) supports solutions for any positive value of $\delta$,
$\delta$ cannot be too large for a hinge soliton solution.
On the one hand, the width of the soliton decreases as $\delta$ increases, meaning that
if $\delta$ is very large, the width is very small so that we can not apply the approximation
that the wave packet varies slowly in space.
On the other hand, for large $\delta$, $\mu=\varepsilon_{k_z}+\delta$
can be moved into the bulk energy spectra, resulting in the scattering of the soliton into the
bulk states.
For a BEC, the total atom number $N$ is given by $\sum_{\lambda}\int|\psi_\lambda(\textbf{r})|^2 d\textbf{r}=N$.
For clarity, if we consider a zero-order approximation based on Eq.~(\ref{Eq_psi_tayl})
so that ${\psi}_\lambda({\bf r},t)\approx e^{-i\varepsilon_{k_z} t+i k_z z} \varphi_{k_z,\lambda} a_B(z,t)$,
the atom number constraint yields $\int |a_B(z,t)|^2 dz =N$. We thus obtain
$\delta=-\frac{g_{eff}^2 N^2}{8\varepsilon''}$, showing that the bright soliton becomes narrower
with its peak value being increased as we increase $g$ with $N$ being fixed.

To show that a bright soliton localized on a hinge can exist in the second-order topological insulator,
we now study the dynamics of the soliton by solving the nonlinear Schr\"odinger equation numerically.
We first construct an initial wave packet using the hinge modes in the strong topological insulator
based on Eqs.~(\ref{Eq_psi_tayl}) and (\ref{Eq_a_bri})
and then observe the dynamics of the wave packet as time evolves.
Fig.~\ref{Fig2}(a) illustrates that
the wave packet moves unidirectionally along the $z$ axis without changing its shape with the velocity determined by $\varepsilon'$,
implying the existence of a travelling soliton.
In comparison, we also calculate the dynamics of the same initial wave packet in the linear case with $g=0$ (without nonlinearities)
and plot the results in Fig.~\ref{Fig2}(b).
The figure shows that the wave packet diffuses significantly as time progresses.
The diffusion of the wave packet happens in the linear scenario because the hinge spectrum is not perfectly linear. To
show the difference quantitatively, we further compare the time evolution of the maximum value and the full width at half maximum of the wave packet for
these two cases in Fig.~\ref{Fig2}(c). We can see clearly that in the nonlinear case, the peak value
becomes stable after some slight oscillations, whereas in the linear case, it suffers a continuous decline.
The propagation of the hinge soliton over time is also visually displayed in Fig.~\ref{Fig2}(d).
The above analysis demonstrates the existence of a stable bright hinge soliton in the second-order topological insulator.

\begin{figure}[t]
\includegraphics[width = 0.99\linewidth]{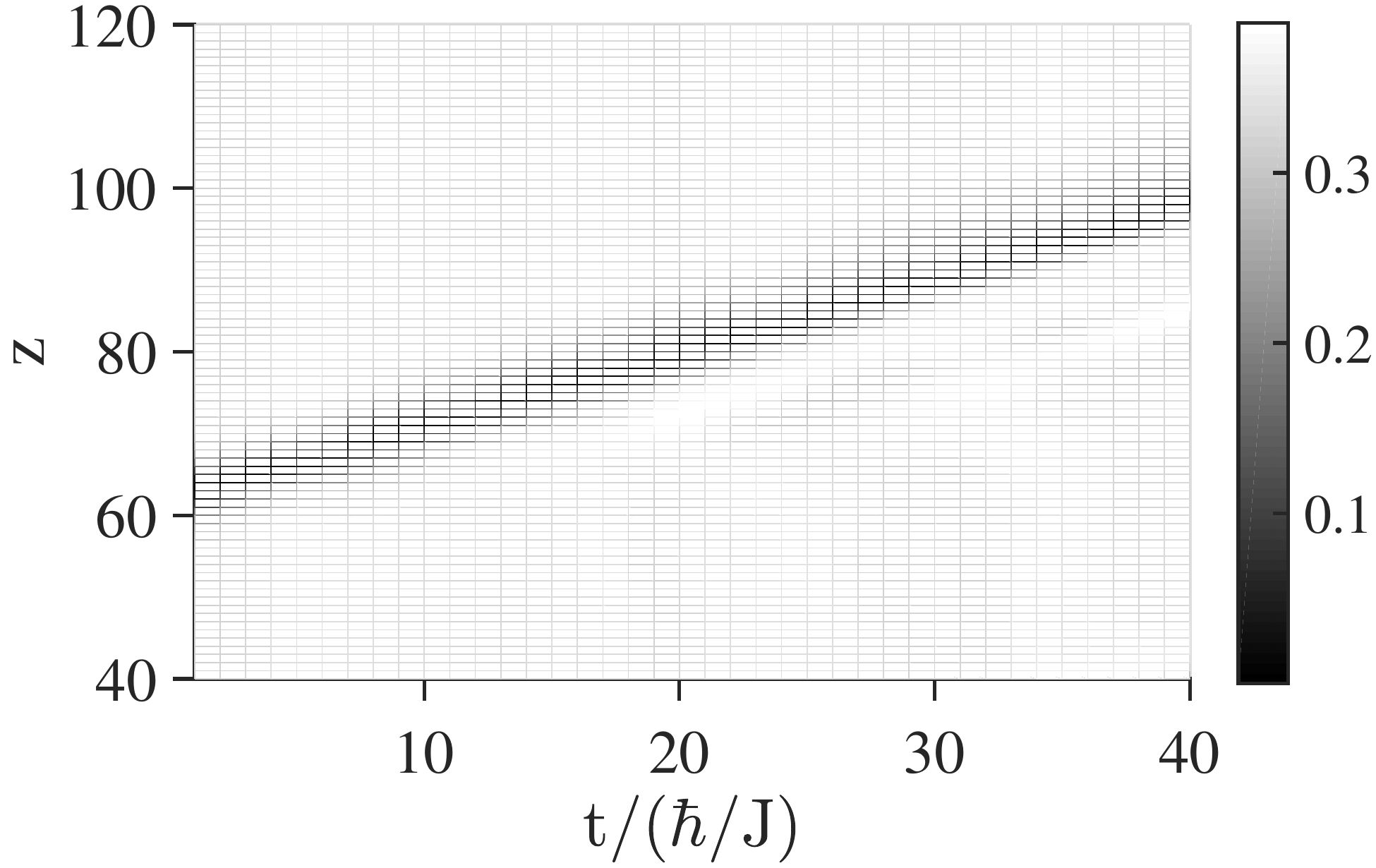}
\caption{The propagation of a wave packet over time, showing the existence of a dark soliton. The wave packet is initialized at $t=0$
based on Eq.~(\ref{Eq_psi_tayl}) and Eq.~(\ref{Eq_a_dark}) with $k_z=6\pi/7$, $\varepsilon''=0.4374$ and $\delta=0.07J$.
Here $M=2J$. }
\label{Fig3}
\end{figure}

When $\varepsilon''>0$ for $g_{\textrm{eff}}>0$, Eq.~(\ref{Eq_equ_a}) allows for the existence of the following dark soliton solution
with a density dip,
\begin{align}\label{Eq_a_dark}
a_{\rm{D}}(z,t)=(\delta/g_{\textrm{eff}})^{1/2}\rm{tanh}[(\delta/\varepsilon'')^{1/2}(z-\varepsilon' t)]e^{-i\delta t}.
\end{align}
For a BEC with a fixed atom number, similar to the case of the bright soliton, if we consider atoms trapped
in a cubic box with size $L\gg 1$, then we obtain an expression of $\delta$ in terms of $g_{eff}$ and $N$, i.e.,
$
\delta=[\frac{\sqrt{\varepsilon''}}{L}+\sqrt{\frac{\varepsilon''}{L^2}+\frac{N g_{eff}}{L}}]^2
$. It indicates that the dark soliton becomes narrower when $g$ is increased.
To numerically demonstrate the dynamics of the dark soliton, we first construct an initial wave packet with a density dip based on Eqs.~(\ref{Eq_psi_tayl})
and (\ref{Eq_a_dark}) and then compute the dynamics of the wave packet. Fig.~\ref{Fig3} illustrates that the wave packet travels along $z$ with its dip shape remaining almost unchanged during the time evolution. This demonstrates that a dark soliton can stably exist
on the hinges in the second-order topological insulator.

\begin{figure}[t]
\includegraphics[width = 0.94\linewidth]{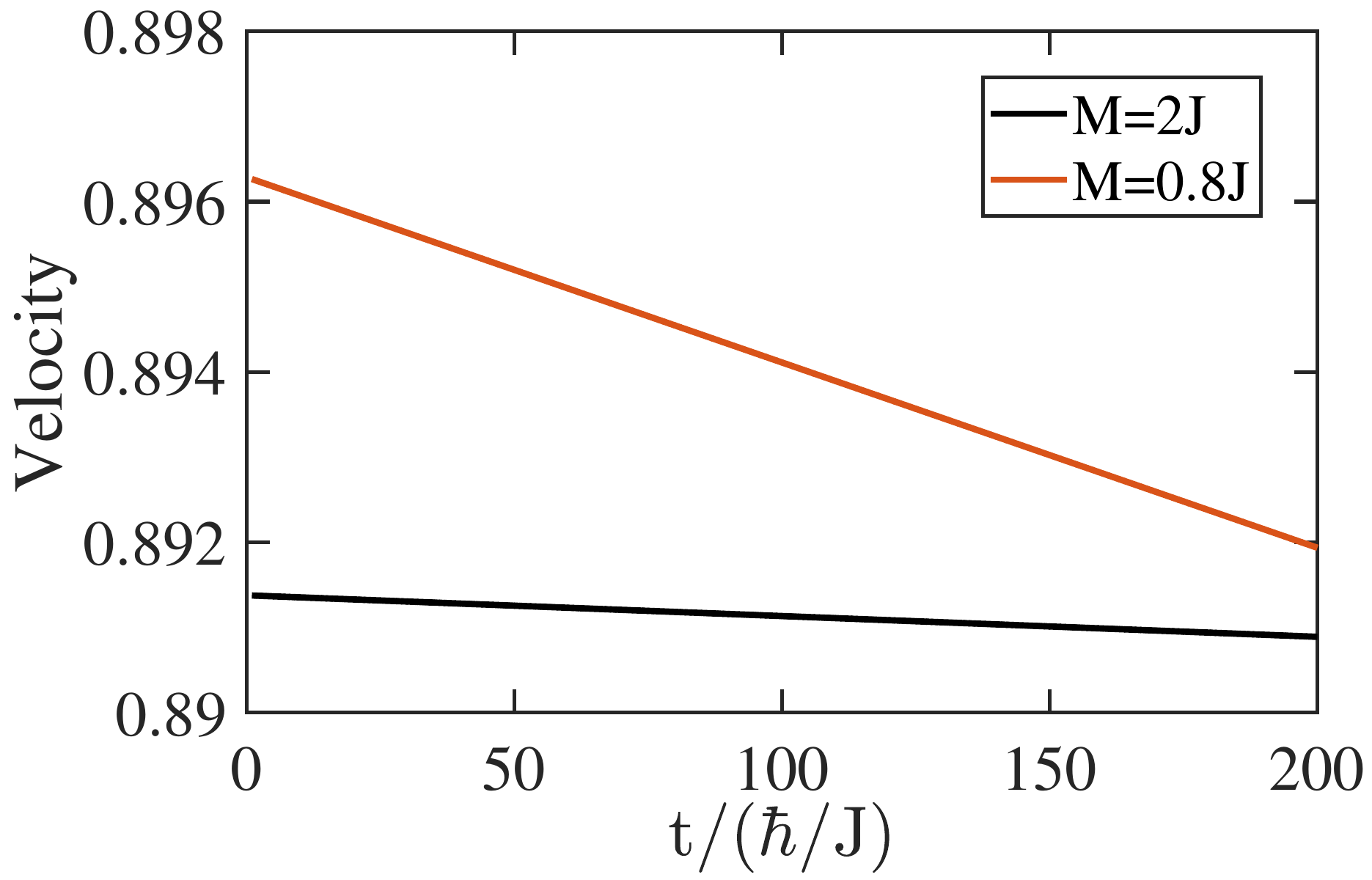}
\caption{The change of a soliton's velocity as time evolves. The soliton is initialized at $t=0$
based on Eq.~(\ref{Eq_psi_tayl}) and Eq.~(\ref{Eq_a_bri}) with $k_z=8\pi/7$ and $\delta=0.007J$.
The black and red lines depict the scenarios with $M=2J$ and $M=0.8J$, respectively.
Here the step size of time is $\Delta t=1\times10^{-3}\hbar/J$.}	
\label{Fig4}
\end{figure}

In the strong topological regime when $1<|M/J|<3$, there are chiral hinge modes in the energy gap. In contrast, in the weak topological regime when $|M/J|<1$, instead of being chiral for the hinge modes, there are two states with opposite group velocities on each hinge for each energy in the energy gap [see Fig.~\ref{Fig1}(b)].
We now apply the same method to explore whether a stable bright soliton can exist on a hinge of a
weak second-order topological insulator.
Similar to the strong topological case, we find that the wave packet travels steadily along $z$ without noticeable change of its shape, suggesting
the existence of a stable bright soliton. But the soliton gradually slows down over time as shown in Fig.~\ref{Fig4}, possibly due to the scattering of the state to other modes with the same energy and opposite
group velocity. In comparison, we also plot the velocity of the soliton with respect to time for the former case
with $M=2J$ in Fig.~\ref{Fig4},
showing that the velocity remains stable. While the figure shows a very small decline of the velocity,
it may be caused by the numerical error, as the decline becomes smaller when we use a smaller step size of time in the computation. Note that this numerical error in the latter case with $M=0.8J$ also exists but is not as noticeable as in the former one.

\section{Hinge solitons in electric CIRCUITs}
In this section, we will introduce a
practical scheme to realize the second-order topological insulator in 3D and demonstrate that a bright soliton for the voltage distribution can exist in the presence of nonlinear inductors.

\begin{figure*}[t]
\centering
\includegraphics[width = \textwidth]{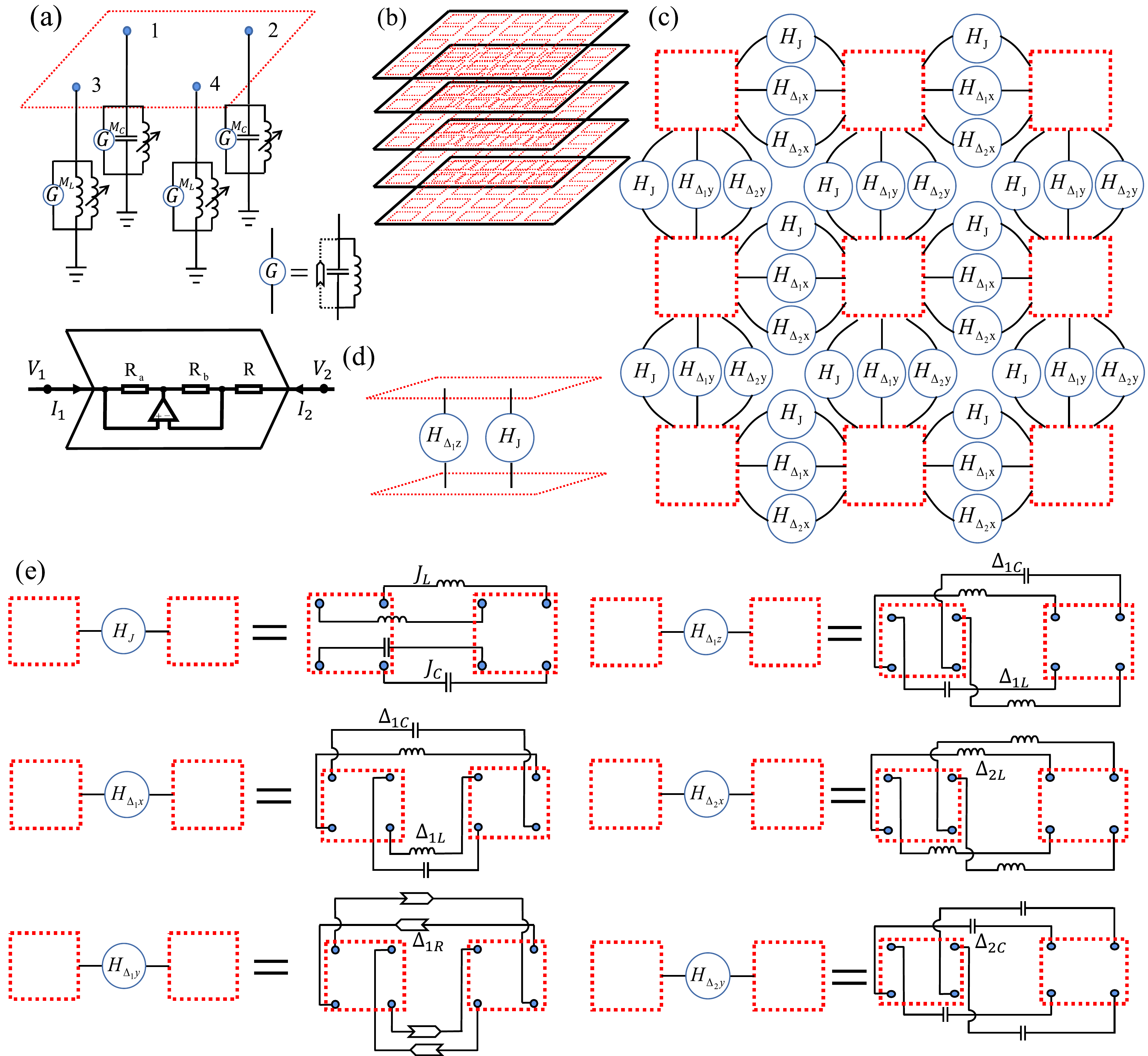}
\caption{(a) Schematics of a unit cell consisting of four nodes, each of which is connected to some grounded electric elements to simulate the diagonal entries
in the Hamiltonian and nonlinear terms in $H_N$. The electric element denoted by a circle labelled by $G$ is used to cancel the
diagonal entries contributed by the electric elements connecting two distinct nodes. The specific configurations for
these elements are listed in Table~\ref{Table1}. The detailed structure of the INIC is displayed in the lower part. (b) Scheme of an electric circuit for simulation of the second-order topological insulator in 3D. The connection inside each layer [in the $(x,y)$ plane] is shown in (c) and the connection between different layers along $z$ is shown in (d). (e) The detailed electric structure for
the circles labelled by $H_J$, $H_{\Delta_1\nu}$ ($\nu=x,y,z$) and $H_{\Delta_2\nu}$ ($\nu=x,y$) in (c) and (d).
$J_L$, $\Delta_{1L}$ and $\Delta_{2L}$ denote the inductances of the corresponding inductors, $J_C$, $\Delta_{1C}$ and $\Delta_{2C}$ denote the capacitances of the corresponding capacitors, and $\Delta_{1R}$
denotes the resistance of the INIC (corresponding to $R$ in the lower part of (a)).}	
\label{Fig5}
\end{figure*}

Let us now consider an electric network with a number of nodes labelled by $m$. Suppose that an electric current $I_m$ is
externally flowing into node $m$ and the current then flows into other nodes, which are connected with node $m$.
According to Kirchhoff's law, we have
\begin{equation}\label{Eq_krich}
{I}_m=\sum_n {I}_{mn}=\sum_{n}Y_{mn}({V}_m-{V}_n)+Y_m {V}_m,
\end{equation}
where ${I}_{mn}$ denotes the current flowing from node $m$ to node $n$, $Y_{mn}$ and $Y_m$ represent the admittance between
node $m$ and node $n$ and the admittance between node $m$ and the ground, respectively, and ${V}_m$ denotes the
voltages measured at node $m$ against ground. The relation above can be written in a matrix form as
\begin{equation}\label{Eq_I_LV}
{I}=L{V},
\end{equation}
where $L$ is the Laplacian with $L_{mn}=-Y_{mn}+\delta_{mn}(Y_m+\sum_{n}Y_{mn})$ and ${I}$ and ${V}$ are the column vectors consisting of electric currents and potentials at each node. Consider the alternating current with the frequency of $\omega_0$ (i.e., $I=\mathcal{I}e^{i\omega_0 t}$
and $V=\mathcal{V}e^{i\omega_0 t}$), we have
\begin{equation}\label{Eq_I_IWLV}
\mathcal{I}=i\omega_0 \mathcal{L}(\omega_0)\mathcal{V}.
\end{equation}

To simulate the Hamiltonian in Eq.~(\ref{EqHLk}), we need to engineer electric connections so that $\mathcal{L}=H_L$ where
$H_L$ is the real space Hamiltonian of $H_L({\bf k})$ in Eq.~(\ref{EqHLk}). Since there are four degrees of freedom in each unit cell,
we consider four nodes in each unit cell as shown in Fig.~\ref{Fig5}(a). Evidently, inductors and capacitors between two neighboring
nodes contribute terms corresponding to the hopping with positive and negative amplitude, respectively. For the
hopping with imaginary amplitudes, we utilize negative impedance converters with current inversion (INICs)~\cite{Hofmann2019PRL}, whose structure is shown in Fig.~\ref{Fig5}(a). Its node voltage equation is
\begin{equation}
\label{Eq_INIC}
\mathcal{I}_N=\frac{1}{R}
\left[
\begin{array}{cc}
-\upsilon&\upsilon\\
-1&1
\end{array}
\right]
\mathcal{V}_N,
\end{equation}
where $\mathcal{I}_N=\left(
             \begin{array}{cc}
               \mathcal{I}_1 & \mathcal{I}_2 \\
             \end{array}
           \right)^T
$, $\mathcal{V}_N=\left(
          \begin{array}{cc}
            \mathcal{V}_1 & \mathcal{V}_2 \\
          \end{array}
        \right)^T
$ and $\upsilon=R_b/R_a$ (consider $\upsilon=1$ here) as shown in Fig.~\ref{Fig5}(a). Clearly, when the current
flows following the direction of the INIC (the large arrow), the resistance is negative and it is positive,
when the direction is opposite.

The scheme shown in Fig.~\ref{Fig5} realizes a Laplacian that simulates the Hamiltonian in Eq.~(\ref{EqHLk})
[i.e., $\mathcal{L}(\omega_0)=H_E(\omega_0)$] with
\begin{eqnarray}\label{Eq_HE}
&&H_E(\omega_0)=(M_E+J_E\sum_{\nu=x,y,z}\cos k_\nu)\tau_z \\
&&+\Delta_{E1}\sum_{\nu=x,y,z}\sin k_\nu\tau_y \sigma_\nu+\Delta_{E2}(\cos k_x-\cos k_y)\tau_x, \nonumber
\end{eqnarray}
when $J_C=J_E/2$, $J_L=2/(\omega_0^2 J_E)$, $\Delta_{1C}=\Delta_{E1}/2$, $\Delta_{1L}=2/(\omega_0^2 \Delta_{E1})$,
$\Delta_{2C}=\Delta_{E2}/2$, $\Delta_{2L}=2/(\omega_0^2 \Delta_{E2})$, and $1/(\omega_0^2 M_L)=M_C=M_E$. This Hamiltonian takes the same form
as Eq.~(\ref{EqHLk}) after a transformation: $\tau_x\rightarrow \tau_y$ and $\tau_y \rightarrow \tau_x$ which doesn't affect topological properties of the system.
For simplicity, we set $\Delta_{E1}=\Delta_{E2}=J_E$.
In the scheme, the connections inside each layer (the connections between distinct layers) are displayed in Fig.~\ref{Fig5}(c)
[Fig.~\ref{Fig5}(d)] with detailed electric
structures shown in Fig.~\ref{Fig5}(e). In each unit cell, there are four nodes connected to the ground with electric elements
contributing to the diagonal terms. The electric element denoted by a circle labelled by $G$ is employed to
cancel the diagonal entries contributed by the electric elements connecting two distinct nodes. Explicit
configurations for the elements are listed in Table~(\ref{Table1}). Additional linear inductors and capacitors are
used to generate the mass term. This electric circuit can be utilized to study the topological properties of the second-order topological insulator in 3D if the nonlinear inductor is not involved.

\begin{table}[t]
	\centering
	\caption{Configurations for the electric element denoted by a circle labelled by $G$ in Fig.~\ref{Fig5}(a) for different positions for a system in a geometry with open boundaries
in the $(x,y)$ plane with $N_x\times N_y$ unit cells and periodic boundaries along $z$. The units of capacitors, inductors
and INICs are $J_E$, $1/(\omega_0^2 J_E)$ and $1/(\omega_0 J_E)$, respectively. The arrow indicates the direction of
the INICs and the symbol $/$ indicates that the INICs are not required.}
	\begin{tabular}{cccc}
		\toprule
		\multirow{2}*{Position}  & \multicolumn{3}{c}{Elements}   \\
		\cline{2-4}
		&    Capacitor & Inductor & INICs \\
		\midrule
		Bulk &  ($6$,$6$,$3$,$3$) & ($\frac{1}{3}$,$\frac{1}{3}$,$\frac{1}{6}$,$\frac{1}{6}$)  &(/,/,/,/) \\
		$(x,y)=(1,1)$ &  ($6.5$,$6.5$,$5$,$5$) & ($\frac{1}{5}$,$\frac{1}{5}$,$\frac{1}{6.5}$,$\frac{1}{6.5})$  &($\downarrow$,$\uparrow$,$\uparrow$,$\downarrow$) \\
		$(x,y)=(N_x,1)$ &  ($7$,$7$,$4.5$,$4.5$) & ($\frac{1}{4.5}$,$\frac{1}{4.5}$,$\frac{1}{7}$,$\frac{1}{7}$)  &($\downarrow$,$\uparrow$,$\uparrow$,$\downarrow$) \\
		$(x,y)=(1,N_y)$ &  ($6.5$,$6.5$,$5$,$5$) & ($\frac{1}{5}$,$\frac{1}{5}$,$\frac{1}{6.5}$,$\frac{1}{6.5}$)  &($\uparrow$,$\downarrow$,$\downarrow$,$\uparrow$) \\
		$(x,y)=(N_x,N_y)$ &  ($7$,$7$,$4.5$,$4.5$) & ($\frac{1}{4.5}$,$\frac{1}{4.5}$,$\frac{1}{7}$,$\frac{1}{7}$)  &($\uparrow$,$\downarrow$,$\downarrow$,$\uparrow$) \\
		$\text{Surface}({y=1})$ &  ($6.5$,$6.5$,$4$,$4$) & ($\frac{1}{3.5}$,$\frac{1}{3.5}$,$\frac{1}{6}$,$\frac{1}{6}$)  &($\downarrow$,$\uparrow$,$\uparrow$,$\downarrow$) \\
		$\text{Surface}({y=N_y})$ &  ($6.5$,$6.5$,$4$,$4$) & ($\frac{1}{3.5}$,$\frac{1}{3.5}$,$\frac{1}{6}$,$\frac{1}{6}$)  &($\uparrow$,$\downarrow$,$\downarrow$,$\uparrow$) \\
		$\text{Surface}({x=1})$ &  ($6$,$6$,$4$,$4$) & ($\frac{1}{4.5}$,$\frac{1}{4.5}$,$\frac{1}{6.5}$,$\frac{1}{6.5}$)  &(/,/,/,/) \\
		$\text{Surface}({x=N_x})$ &  ($6.5$,$6.5$,$3.5$,$3.5$) & ($\frac{1}{4}$,$\frac{1}{4}$,$\frac{1}{7}$,$\frac{1}{7}$)  &(/,/,/,/) \\
		\bottomrule
	\end{tabular}
	\label{Table1}
\end{table}

\begin{figure*}[t]
\centering
\includegraphics[width = 1\linewidth]{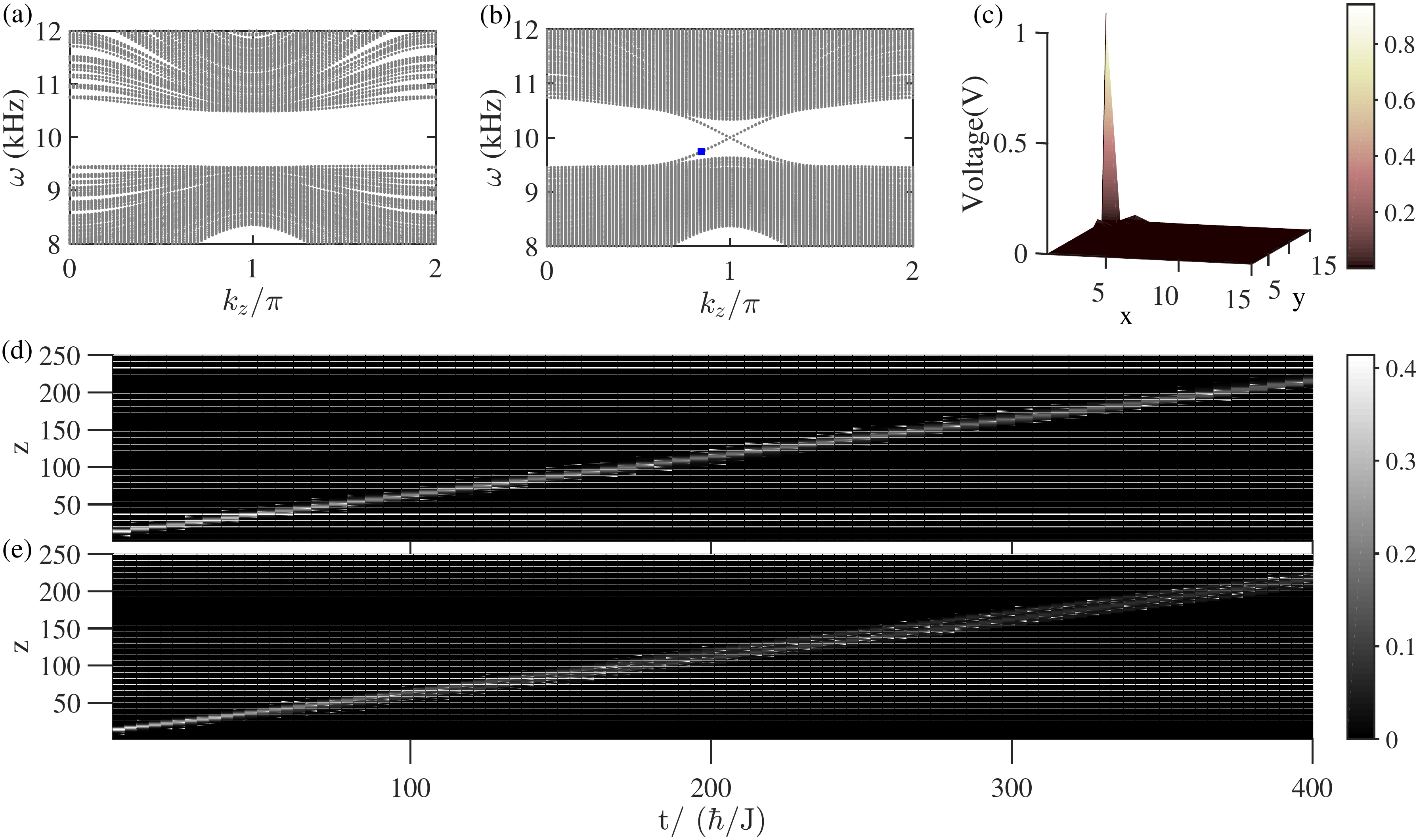}
\caption{The eigenfrequency spectra versus momentum $k_z$ of $H_{\textrm{HTIC}}$ with (a) periodic boundaries and (b) open boundaries
along $x$ and $y$. Note that only positive parts of the spectra are plotted.
The lines traversing the gap describe the hinge-localized states. (c) The voltage distribution of a hinge mode of $H_{\textrm{HTIC}}$ in the $(x,y)$ plane at $k_z=6\pi/7$,
described by $\sqrt{\sum_{\lambda=5,6,7,8} \phi_{k_z,\lambda}(x,y)^2}$,
where $\phi_{k_z}(x,y)$ is the eigenvector of $H_{\textrm{HTIC}}$ corresponding to the eigenfrequency labelled by the blue squares in (b)
of $H_{\textrm{HTIC}}$
with open boundaries along $x$ and $y$ and periodic boundary along $z$.
The time evolution of a voltage pulse characterized by $\sqrt{\sum_{\lambda=5,6,7,8}|\Psi_{\lambda}(t)|^2}$ for (d) $g_c=10\textrm{mH}^{-1}V^{-2}$ and (e) $g_c=0$. The voltage pulse is initialized
at $t=0$ to the state described by Eq.~(\ref{EqIn}) with $a_0=1$, $\lambda_0=0.5$ and $k_z=6\pi/7$.
Here
$\Delta_{1C}=\Delta_{2C}=J_C=50\, \mu\textrm{F}$, $\Delta_{1L}=\Delta_{2L}=J_L=200\, \mu\textrm{H}$, $\Delta_{2R}=2\,\Omega$, and $M_C=100\,\mu\textrm{F}$.
}
\label{Fig6}
\end{figure*}

To simulate the dynamics of a soliton, we solve the following equation of motion of the circuit~\cite{Hofmann2019PRL}
\begin{align}\label{CircuitD}
\frac{d}{dt}{I}(t)=\mathcal{C}\frac{d^2}{dt^2}{V}(t)+\Sigma\frac{d}{dt}{V}(t)+\Pi{V}(t)
\end{align}
where $\mathcal{C}$, $\Sigma$, and $\Pi$ are the real-valued matrices
characterizing the capacitance of capacitors, the conductance of the INICs, and the inductance of inductors, respectively..
These matrices in momentum space are given by
\begin{eqnarray}\label{EqLRC_1}
\mathcal{C}(\bm{k})&=&A_C+M \tau_z-2J_C\sum_{\nu=x,y,z}\cos k_\nu(\tau_0-\tau_z)/2 \nonumber \\
&&-2\Delta_{2C}\cos k_y \tau_x-\Delta_{1C}\cos k_x\tau_x\sigma_x -\Delta_{1C}\cos k_z\tau_x \nonumber \\
&&+\Delta_{1C}\sin k_x\tau_y\sigma_x+\Delta_{1C}\sin k_z \tau_y\sigma_z\\
\Sigma(\bm{k})&=&\frac{2}{\Delta_{1R}}\sin k_y \tau_y\sigma_y
\end{eqnarray}
\begin{eqnarray}\label{EqLRC_2}
	\Pi(\bm{k})&=&A_L-\frac{2}{J_L}\sum_{\nu=x,y,z}\cos k_\nu(\tau_0+\tau_z)/2 \nonumber\\
	&&-\frac{2}{\Delta_{2L}}\cos k_x \tau_x
	-\frac{1}{\Delta_{1L}}(\cos k_x \tau_x\sigma_x+\sin k_x \tau_y\sigma_x) \nonumber \\
	&&-\frac{1}{\Delta_{1L}}(\cos k_z \tau_x+\sin k_z \tau_y \sigma_z),
\end{eqnarray}
where $A_C+M\tau_z$ and $A_L$ are contributed by the grounded capacitors and inductors, respectively.
When $J_{C}=\Delta_{1C}$ and $J_L=\Delta_{1L}$ without $M_C$ and $M_L$, we have $A_C=8J_{C}$, $M=2J_C$,
and $A_L=8/\Delta_{1L}$ using the grounding configurations listed in Table~\ref{Table1}.

For alternating input currents with the frequency of $\omega$ [$I(t)=\mathcal{I}e^{i\omega t}$ and $V(t)=\mathcal{V}e^{i\omega t}$], this differential equation
gives rise to the circuit Laplacian
\begin{align}\label{Eq_Lapla}
L(\omega)=i\omega \mathcal{C}+\Sigma+\frac{1}{i\omega}\Pi=i\omega \mathcal{L}(\omega)
\end{align}
so that $\mathcal{I}=L(\omega)\mathcal{V}$, which is the same as Eq.~(\ref{Eq_I_IWLV}). 

Let us now consider a specific case without any input currents, i.e., $I(t)=0$. In this
case, we reduce the Eq.~(\ref{CircuitD}) to the following first-order
differential equation similar to the Schr\"odinger equation for the higher-order topological insulator circuit (HTIC),
\begin{align}\label{Eq_Sch_H_C}
-i\frac{d}{dt}\Psi={H}_{\textrm{HTIC}}\Psi,
\end{align}
where $\Psi=(\dot{{V}}(t),{V}(t))^T$, and
\begin{align}\label{Eq_H_HTIC}
{H}_{\textrm{HTIC}}=i
\left[
\begin{array}{cc}
\mathcal{C}^{-1}\Sigma&\mathcal{C}^{-1}\Pi\\
-\mathds{1}&0
\end{array}
\right ].
\end{align}
This differential equation has the solution of $\Psi(t)=e^{i\omega_n t}\phi_n$, where $\phi_n$ is an eigenvector of
$H_{\textrm{HTIC}}$ corresponding to the eigenvalue $\omega_n$.
Clearly, the eigenvalues of $H_{\textrm{HTIC}}$ appear in pairs
as $(\omega_n,-\omega_n^*)$.
While $H_{\textrm{HTIC}}$ is non-Hermitian, its eigenvalues are real due to
the positive semidefinite properties of $\mathcal{C}$ and $\Pi$~\cite{Hofmann2019PRL}, implying that the eigenvalues
emerge in pairs as $(\omega_n,-\omega_n)$.

Since $\omega_n$ is the alternating current frequency allowed by Eq.~(\ref{Eq_Sch_H_C}),
we must have $L(\omega_n)\mathcal{V}=0$ given that the input currents are zero.
To have nontrivial solutions, we require $\det(L(\omega_n))=0$. In other words, there exist eigenvectors $V_n$ of $L(\omega_n)$ with zero eigenvalue,
i.e.,
\begin{equation}
L(\omega_n)V_n=0.
\end{equation}
One can change the equation into the following form
\begin{equation}
H_{\mathrm{HTIC}}\phi_n=\omega_n \phi_n,
\end{equation}
where
\begin{equation}
\phi_n=(\begin{array}{cc}
           i\omega_n V_n & V_n
         \end{array}
)^T.
\label{PhiVn}
\end{equation}
It also tells us that if $L(\omega_n)$ has zero eigenvalues, then $\omega_n$ is an eigenvalue of $H_{\textrm{HTIC}}$.

The intimate connection suggests that the band structures of $H_{\textrm{HTIC}}$ and $H_E(\omega_0)=\mathcal{L}(\omega_0)$
share some common properties.
For example, if $H_E(\omega_0)$ is in a higher-order topological phase with chiral hinge
modes, then there exist four zero-energy hinge modes at $k_z=\pi$ [see Fig.~\ref{Fig1}(a)].
Based on the discussion above, $\omega_0$ should be an eigenvalue of $H_{\textrm{HTIC}}$.
In addition, according to Eq.~(\ref{PhiVn}),
we have
four modes of $H_{\textrm{HTIC}}$ with frequency of $\omega_0$ localized on the hinges,
suggesting that $H_{\textrm{HTIC}}$ has chiral hinge modes.
Indeed, Fig.~\ref{Fig6}(b) illustrates that $H_{\textrm{HTIC}}$ exhibits the chiral hinge modes
across the gap, which do not exist for periodic boundaries along $x$ and $y$ [see Fig.~\ref{Fig6}(a)].
The voltage distribution of these states also shows that they are localized on the hinges [see Fig.~\ref{Fig6}(c)].
We note that gapless nontrivial states in circuits have also been found in other 2D systems~\cite{Hofmann2019PRL,Qibo2020prb}.

In stark contrast to a soliton for the atomic density in BECs, in our case, a soliton is for the voltage distribution in the
electric circuit. To generate the soliton, we add nonlinear inductors in the circuit, which may be realized by a voltage-controlled variable inductor~\cite{MTonso2005,Liu2012}. Each nonlinear inductor connects each node with the ground (see Fig.~\ref{Fig5}), contributing a nonlinear
inductance described by
\begin{equation}\label{Eq_non_L}
\Pi_{N({\bf r},\lambda),({\bf r}^\prime,\lambda^\prime)}=g_c\delta_{{\bf r}{\bf r}^\prime}\delta_{\lambda\lambda^\prime}|V_{{\bf r},\lambda}|^2,
\end{equation}
where $V_{{\bf r},\lambda}$ denotes the voltage at the node labelled by ${\bf r}, \lambda$.

To create a bright soliton, we initialize the state at $t=0$ as
\begin{equation} \label{EqIn}
\Psi({\bf r},t=0)=a_0 {\textrm{sech}}(\lambda_0 z)e^{ik_z z}\phi_{k_z}(x,y),
\end{equation}
and then observe its dynamics. Here $a_0$ and $\lambda_0$ are real parameters and $\phi_{k_z}(x,y)$ is a hinge mode of $H_{\textrm{HTIC}}$ for a
quasimomentum of $k_z$
in a geometry with open boundaries along $x$ and $y$
and periodic boundary along $z$.
To simulate the dynamics, we impose open boundary conditions along all directions, in which case
the parameters of the grounding elements at the corners
and on the surface perpendicular to the $z$ axis need to be adjusted to cancel the diagonal parts
contributed by the electric elements connecting distinct nodes. For a specific example with $g_c=10\textrm{mH}^{-1}V^{-2}, k_z=6\pi/7$,
$\Delta_{1C}=\Delta_{2C}=J_C=50\, \mu\textrm{F}$, $\Delta_{1L}=\Delta_{2L}=J_L=200\, \mu\textrm{H}$, $\Delta_{2R}=2\,\Omega$, and $M_C=100\,\mu\textrm{F}$, we take
$a_0=1$ and $\lambda_0=0.5$. In Fig.~\ref{Fig6}(d) and (e),
we plot the dynamics of the voltage peak with and without nonlinearity, respectively.
It can be seen that the voltage distribution with nonlinearity is stable without conspicuous
dispersion as time evolves. In stark contrast, in the case without nonlinearity, the wave packet
for the voltage distribution becomes wider as it evolves over time. This shows the existence of a hinge bright soliton in
the nonlinear circuit system.

\section{CONCLUSION}
In summary, we have theoretically studied solitons in a 3D second-order topological insulator when nonlinearity is involved.
We find that solitons can exist in both the strong and weak regimes, but the soliton in the former regime is more stable than that in the latter regime given that the
latter one gradually slows down as time progresses.
In addition, the soliton on a hinge in the strong regime can only move along one direction due to the chiral property of hinge modes.
To realize the soliton in a practical experimental system, we further introduce an electric network to simulate the 3D second-order topological insulator.
By calculating the dynamics of a voltage pulse in the electric circuit, we show the existence of a soliton for the voltage distribution when nonlinear inductors are involved.
Such solitons can also be experimentally observed in other nonlinear systems, such as ultracold atomic gases and nonlinear optics.

\begin{acknowledgments}
This work is
supported by the start-up fund from Tsinghua University,
the National Thousand-Young-Talents Program and the National Natural Science Foundation
of China (11974201).
\end{acknowledgments}

\end{document}